\documentclass{article}
\usepackage{graphicx}
\usepackage[breaklinks]{hyperref}
\hypersetup{colorlinks,urlcolor=black,citecolor=black,linkcolor=black,filecolor=black}
\usepackage{breakurl}

\AtBeginDocument{}

\def\sig{{\cal I}}

\begin{document}
\title{Optimal one-parameter observables for the Abeliaqn $Z'$ in
$e^+e^-\to\mu^+\mu^-$ process}

\author{ALEXEY GULOV\\[10pt]
{\it Dnipropetrovsk National University, Dnipropetrovsk, Ukraine}}

\maketitle

\begin{abstract} 
To detect off-shell Abelian $Z'$ boson in $e^+e^-\to\mu^+\mu^-$, we propose one-parameter observables with the best value-to-uncertainty ratio. The observables are constructed by angular integration of the differential cross section with smooth weight functions. The value-to-uncertainty ratio is used as a criterion to select the unique weight function leading to the observable with the best statistical power for data analysis. The observables allow to select either vector or axial-vector $Z'$ couplings to leptons. The obtained observable can be useful in future experiments at lepton colliders such as the ILC.
\end{abstract}

\section{Introduction}

The advanced electron-positron collider ILC is actively discussed in the literature as an important component of future experiments in high energy physics \cite{Baer:2013cma}. In particular, it may provide opportunity to elucidate blind-spots after the LHC experiments.

An intriguing question of modern elementary particle physics is whether new particles beyond the standard model (SM) exist at energies of TeVs. It is common hope that the LHC will be able to catch such particles as resonances, but a lot of details will remain unknown. Precise measurements at the ILC will allow to estimate different features of new heavy particles. Assuming collision energies $\sqrt{s}=0.25,0.5,1$ TeV, one can conclude that the experiment will probably encounter off-shell physics related to new heavy particles. In this regard, it is important to design effective off-shell observables sensitive to new heavy particles beyond the SM.

In the present paper we investigate possibility to detect signals of the Abelian $Z'$ boson \cite{Leike:1998wr,Langacker:2008yv} in $e^+e^-\to\mu^+\mu^-$ process. This process with quite simple $s$-channel kinematics provides an excellent possibility to test new ideas and approaches. Recently, we described observables to detect the Abelian $Z'$ boson in leptonic processes \cite{Gulov:2010zq}. The observables were mainly applied to data collected in LEP experiments leading to some hints of the particle. In case of $s$-channel process, the observable was constructed as a generalized forward-backward cross section. However, the question about maximal statistical efficiency of observables designed to amplify $Z'$ signals has been left without an answer. In the present paper we try to answer the question.

It is a known fact that the differential cross section for the process of interest with spin-one exchange and massless fermions is a linear combination of only two polynomials. For instance, these polynomials can be expressed through the first three Legendre polynomials. The polynomials form a complete orthogonal system of functions for any quantity constructed by integration of differential cross section over scattering angle. Thus, in integrated cross sections only three lowest order Legendre polynomials can be used as non-trivial weights leading to the `moments' considered in details in \cite{Rizzo:2002pc} in application to spin identification problem. Higher order Legendre polynomials disappear in integration of the cross section leading to some loss of integrated data, if they are used. So, one could ignore higher order polynomials to discuss just mean values of integrated cross sections, constructing various observables to amplify $Z'$ signals.

However, the statistical efficiency of any observable is determined by interplay of the mean value and the variance (we use popular notations `statistical uncertainty' or `statistical error' to refer to the square root of the variance).  Fitting experimental data, the mean value of some signal is to be compared with the statistical uncertainty of the background.  The signal under consideration is usually small deviation of cross section from the SM value due to new physics, whereas the background consists mainly of the SM contribution with the proper systematic and statistical uncertainties. In case of good quality of data, the signal of new physics must be comparable with the statistical uncertainty. The ratio of the mean value of the signal to the uncertainty is, probably, the most popular characteristic of the signal strength in this case (the so called number of standard deviations, `sigmas' of the signal). In particular, it defines the discovery reach of the signal. In what follows we will call this ratio as the value-to-uncertainty (or signal-to-uncertainty) ratio.

It is a curious fact that the signal-to-uncertainty ratio, being a popular estimator of strategies to search for signals of new physics, is not used as a direct optimization criterion. In the present paper we propose an approach based on maximal signal-to-uncertainty ratio instead of methods restricted by the analysis of the mean values of observables only. In usual experiments, the variance is proportional to the cross section. This feature is a consequence of the Poisson distribution of the number of detected events. However, when the cross sections in different bins are summed altogether into some mean value, the corresponding variances are summed with the squared coefficients. If one uses a Legendre polynomial to integrate the cross section, a wider sequence of other polynomials appears to integrate the variance. As a consequence, to compare the mean value with the statistical uncertainty the basis of the first three Legendre polynomials becomes incomplete and insufficient. Thus, the integration scheme with the best signal-to-uncertainty ratio is a problem beyond the polynomials which can be seen in the cross section itself. Roughly speaking, we can lose some part of the cross section during angular integration with the higher order Legendre polynomials, if this part produces more uncertainty with respect to the observed signal than contributes to the signal. On the other hand, this scheme amplifies other parts of the cross section with better relative uncertainties. The resulting `receipt' to mix angular intervals is beyond the initial few polynomials in the cross section being some smooth weight function. We believe this method is an interesting alternative competing popular strategies of searching for $Z'$ signals in experiments.

Our analysis is based on the following ideas. We assume off-shell $Z'$ state in the decoupling region, since the current low limits on the $Z'$ mass are about 2 TeV \cite{Hod:2013cba,Chatrchyan:2012oaa}. We also take into account relations between the Abelian $Z'$ couplings to the SM particles \cite{Gulov:2000eh}. In order to construct one-parameter observable we propose integration of the differential cross section with a smooth weight function, generalizing the idea of obtaining the total, forward-backward, center-edge and other known integrated cross sections. Numerically, the weight function can be decomposed by the convenient (infinite) basis of orthogonal polynomials inspired by the kinematics of the process, but special choice of the polynomials is not essential for the results at all, being rather a computational trick. We use the value-to-uncertainty ratio for the observable as an indicator of the statistical power of the $Z'$ signal. Being the optimization criterion, the value-to-uncertainty ratio selects uniquely the most powerful one-parameter integrated cross sections to detect  signals of the Abelian $Z'$ boson at the ILC.

\section{The Abelian $Z'$ boson at low energies}
The usual phenomenological description of the $Z'$ boson introduces couplings to the vector and axial-vector currents of the SM fermions as well as the $Z$--$Z'$ mixing. The corresponding Lagrangian can be written as
\begin{eqnarray}
{\cal L}_{\bar{f}fZ}&=&
\frac{1}{2}Z_\mu \bar{f}\gamma^\mu\left[
(v_{fZ}^{\mathrm{SM}}+\gamma^{5}a_{fZ}^{\mathrm{SM}})\cos\theta_{0}
+(v_{f}+\gamma^{5}a_{f})\sin\theta_{0}
\right]f,
\nonumber \\
{\cal L}_{\bar{f}fZ'}&=&
\frac{1}{2}Z'_{\mu}\bar{f}\gamma^{\mu}\left[
(v_{f}+\gamma^{5}a_{f})\cos\theta_{0}
-(v_{fZ}^{\mathrm{SM}}+\gamma^{5}a_{fZ}^{\mathrm{SM}})\sin\theta_{0}
\right]f.
\label{lagr}
\end{eqnarray}
The $Z'$ couplings to the SM scalar fields can be also introduced in a phenomenological manner by adding new terms into covariant derivatives, such couplings determine the $Z$--$Z'$ mixing angle. Sometimes the effective operators of dimension higher than four are included in consideration. However, the corresponding couplings are naturally suppressed being generated either by loop corrections or next-to-leading order terms in inverse heavy mass scales. We ignore such additional $Z'$ interactions in comparison with the leading-order couplings (\ref{lagr}).

As for the Abelian $Z'$ boson associated with the effective $U(1)$ gauge symmetry at low energies, some couplings have to be inter-related. If we consider the single $Z'$ state with the mass of order TeVs, the following relations arise \cite{Gulov:2000eh}
\begin{eqnarray}
&\displaystyle v_{f,T_3=1/2}=v_{f,T_3=-1/2}-2a,
\quad
a_{f,T_3=-1/2}=-a_{f,T_3=1/2}=a,&
\nonumber \\
&\displaystyle \theta_0=-a\frac{\sin(2\theta_W)}{\sqrt{4\pi\alpha_\mathrm{em}}}\left(\frac{m_Z}{m_{Z'}}\right)^2,&
\label{relations}
\end{eqnarray}
where $T_3$ is the third component of the weak isospin, and the fermions are supposed to belong to the same SM doublet.
The detailed motivation of the relations can be found in \cite{Gulov:2000eh}. Let us note that Eqs.~(\ref{relations}) cover a subset of popular $Z'$ models considered under the traditional model-dependent analysis \cite{Baer:2013cma}. Among the models, the LRS and $\chi$ model satisfy the relations. Other models (ALR, SSM, $\eta$, $\psi$) violate some of the conditions beyond Eqs.~(\ref{relations}): they can contain mixed states of several $U(1)$ gauge bosons, more complicated non-Abelian groups and so on. The $Z'$ couplings to the SM fermions in popular models are listed, in particular, in \cite{Osland:2009tn}.

It is worth to note that in case of the lepton universality, the obtained results can be applied even beyond the mentioned subset of the popular models. This is due to the fact, that the process contains only the charged leptons up to the loop corrections. Indeed, the relations $a_e=a_\mu$, $v_e=v_\mu$ do not contradict (\ref{relations}). Also, at high energies far away from the $Z$ peak, the role of the $Z$-$Z'$ mixing becomes negligibly small. Thus, any model with lepton universality and negligibly small $Z$-$Z'$ mixing angle can be considered beyond our results at $\sqrt{s}\ge$ 0.5 TeV up to the loop corrections.

The virtual $Z'$ boson state contributes to the differential cross section of  $e^+e^-\to \mu^+\mu^-$.
In the lowest order in the inverse $Z'$ mass the cross section deviates from its SM value as
\begin{equation}
\frac{d\sigma}{dz}-\frac{d\sigma^\mathrm{SM}}{dz}=
\frac{m^2_{Z}}{4\pi m^2_{Z'}}\left[
F_{1}{a}^{2}+F_{2}{a}{v}_{e}
+F_{3}{a}{v}_{\mu}+F_{4}{v}_{e}{v}_{\mu}+\ldots\right],\label{dsigma}
\end{equation}
where $z=\cos\theta$ is the cosine of the scattering angle and dots stand for higher corrections in the inverse $Z'$ mass.
Factors $F_{i}(\sqrt{s},z)$ arise from the interference between the SM scattering amplitude (including photon and $Z$-boson exchange) and the $Z'$ exchange amplitude. They have to be computed numerically taking into account both the tree level contribution and loop corrections.

\begin{figure}
\centerline{\includegraphics[scale=0.6]{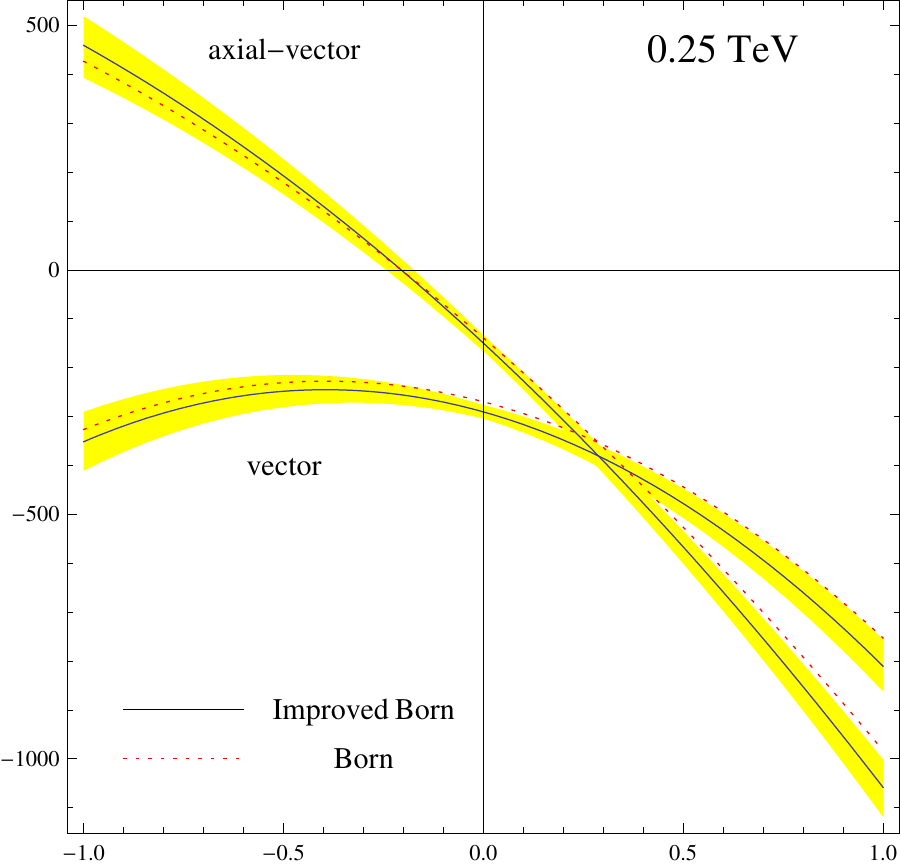}\rule{15pt}{0pt}\includegraphics[scale=0.6]{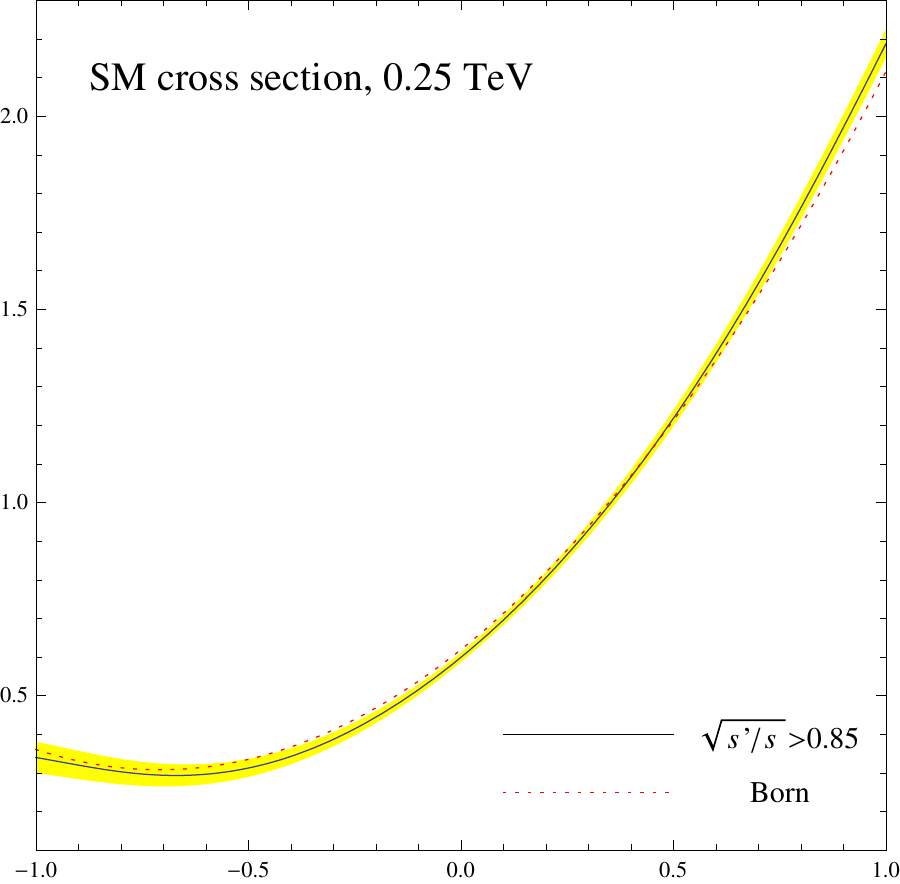}}
\caption{Factors $F_{1}$ (axial-vector), $F_{4}$ (vector), and the SM cross section used in calculations at 250 GeV (in pb). Shaded areas represent possible systematic errors of order $\pm 2\%$ for the SM cross section and $\pm 5\%$ for the factors.  \label{fig:cs250}}
\end{figure}

\begin{figure}
\centerline{\includegraphics[scale=0.6]{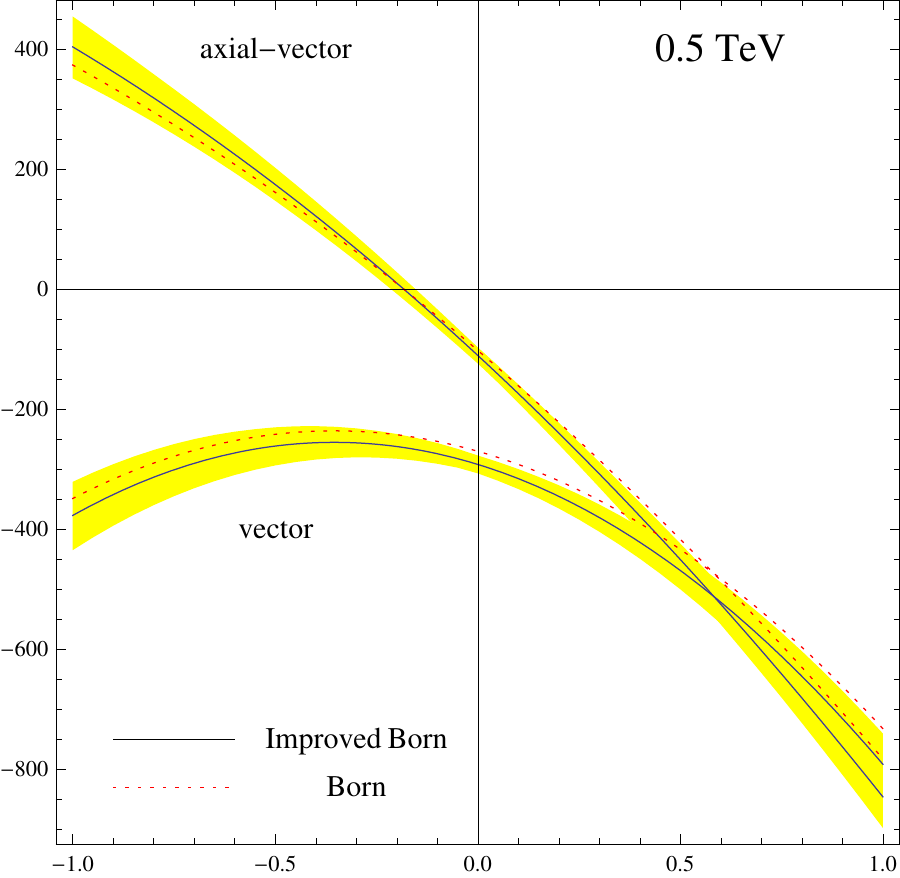}~\includegraphics[scale=0.6]{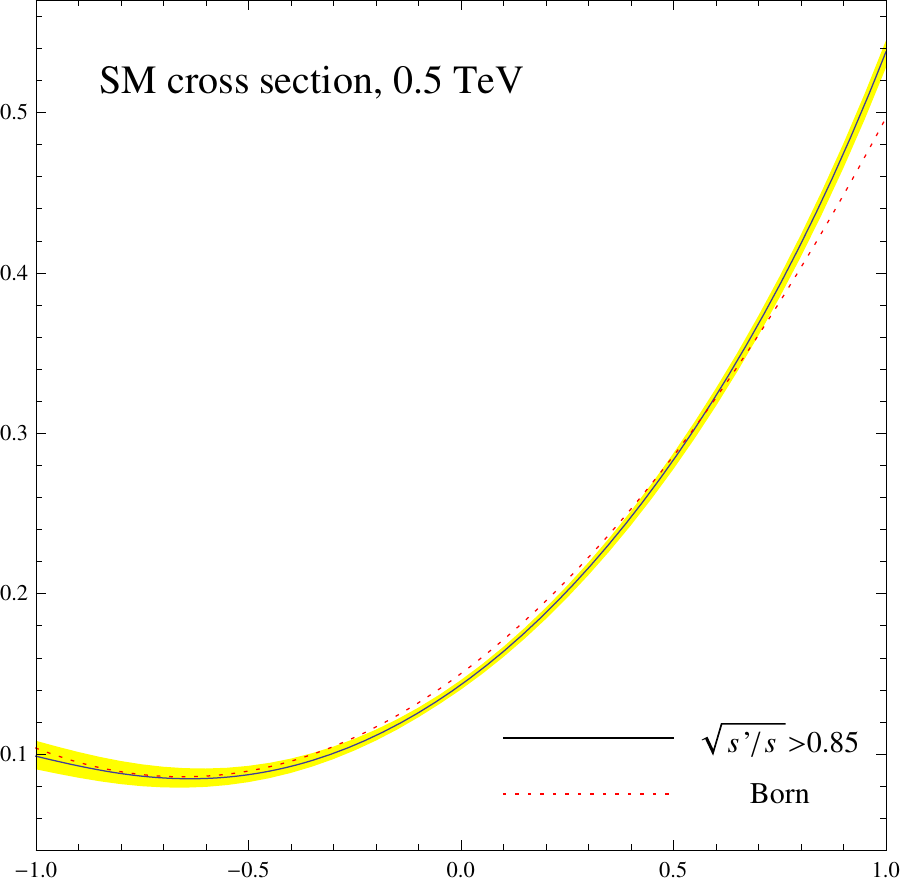}}
\caption{Factors $F_{1}$ (axial-vector), $F_{4}$ (vector), and the SM cross section used in calculations at 500 GeV (in pb). Shaded areas represent possible systematic errors of order $\pm 2\%$ for the SM cross section and $\pm 5\%$ for the factors.  \label{fig:cs500}}
\end{figure}

\begin{figure}
\centerline{\includegraphics[scale=0.6]{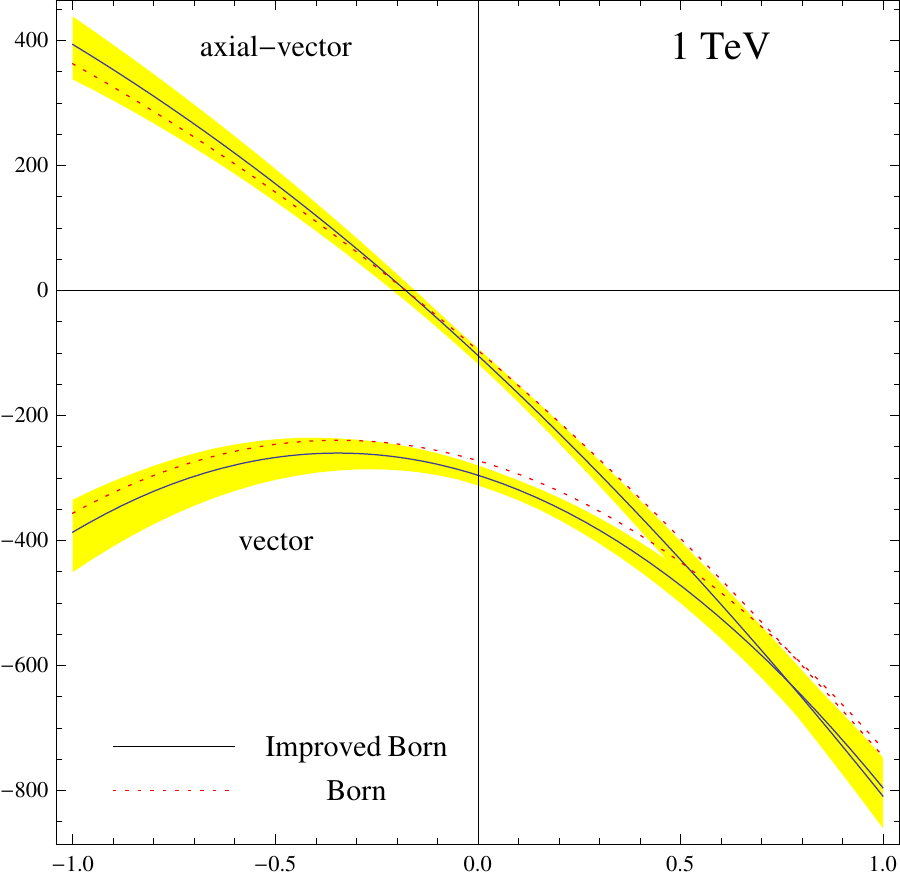}~\includegraphics[scale=0.6]{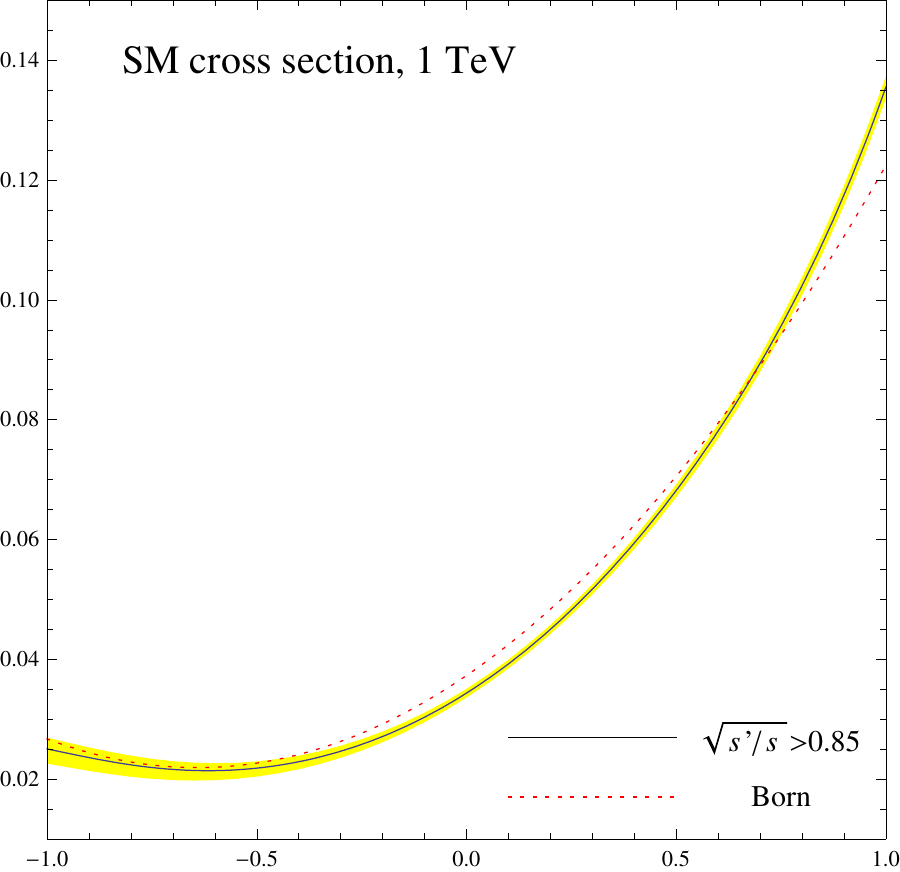}}
\caption{Factors $F_{1}$ (axial-vector), $F_{4}$ (vector), and the SM cross section used in calculations at 1 TeV (in pb). Shaded areas represent possible systematic errors of order $\pm 2\%$ for the SM cross section and $\pm 5\%$ for the factors.  \label{fig:cs1000}}
\end{figure}

Since the complete theory beyond the SM remains unknown, we calculate factors $F_i$ in the improved Born approximation introducing re-summed propagators and running coupling constants. To account for additional effects from box corrections and the soft and hard photon bremsstrahlung we introduce possible systematic error of order $\pm 5\%$. Two leading factors are shown in Figs.  \ref{fig:cs250}-- \ref{fig:cs1000}, where systematic errors correspond to shaded areas around the lines. 

The SM cross section is computed taking into account complete one-loop corrections, the soft and hard photon bremsstrahlung. The calculations are performed with the help of FeynArts \cite{FeynArts}, FormCalc and LoopTools \cite{FormCalc} software. The soft photon bremsstrahlung is included analytically in accordance with \cite{bardin}.  The hard photon bremsstrahlung is accounted with the event selection rule $\sqrt{s'/s}>0.85$, where $s'$ is the Mandelstam variable for the final-state pair.  Integration over $s'$ in the phase space of final particles is performed numerically. Again, the SM cross-sections are shown in Figs.  \ref{fig:cs250}-- \ref{fig:cs1000}. The actual experiment could require additional details and settings such as more complicated integration region in the phase space of final particles, background from other scattering channels and so on. 
In this regard we also introduce possible systematic error of the SM cross-section of order 2\% shown in Figs.  \ref{fig:cs250}-- \ref{fig:cs1000} by shaded areas. 
In principle, the introduced systematic errors can be reduced by more complicated computing tools. In the present paper they are given to estimate the stability of our results with respect to them.

Being measured in experiments, the cross section (\ref{dsigma}) allows to estimate the $Z'$ couplings  ${a}/m_{Z'}$, ${v}_{e}/m_{Z'}$, and ${v}_{\mu}/m_{Z'}$. A non-zero value of some coupling mentioned can be called the $Z'$ signal.

Fitting data, it is better to deal with the minimal number of unknown parameters. One-parameter observables are the most prominent from the statistical point of view. Moreover, a sign-definite observable is more informative, since it can also reject the hypothesis, whereas sign-indefinite observables can only accept the signal. These properties are especially important in case of statistics which is not rich enough to detect signals at high confidence levels. Fortunately, the cross section (\ref{dsigma}) contains one sign-definite term with ${a}^{2}$. If we could select this term in the cross section, we would obtain a powerful observable to detect $Z'$ signals in experiments. In case of lepton universality the term with ${v}_{e}{v}_{\mu}$ also becomes sign-definite. We consider the observables to select either ${a}^{2}$ or ${v}_{e}{v}_{\mu}$ for completeness.

It is also worth to note, that $|F_{2,3}|\ll |F_{1,4}|$. These small factors contribute about 1\% of the resulting sum, and their existence does not affect the key ideas of the present investigation. So, the $Z'$ signal in $e^+e^-\to \mu^+\mu^-$ can be discussed as two-parametric.

\section{Definition of the observable}

The differential cross section (\ref{dsigma}) contains two leading terms at ${a}^{2}$ and ${v}_{e}{v}_{\mu}$. The corresponding factors $F_i(\sqrt{s},z)$ are the functions of energy and scattering angle. We can use angular integration in order to suppress one factor comparing to another. Actually, this means that we will construct some integrated cross section with specific properties.

In general, integrated cross sections are well known in the literature. The most popular integration schemes are based on bin summation with equal weights but opposite signs. As examples, we can mention the total cross section, the forward backward cross-section, the center-edge cross section, etc. However, equal weight of bins is just a possible option. The most general integration scheme can be described by weight function $p(z)$:
\begin{equation}\label{observ}
\sig=\int\limits_{-1}^{1}dz\,p\left(z\right)\frac{d\sigma}{dz}.
\end{equation}
In these notations, the popular mentioned cross sections correspond to step-like weight functions. 
To compare different weight functions and their relative efficiency we choose the normalization, which will be defined below.

The $Z'$ signal is defined as $\sig-\sig^\mathrm{SM}$. $Z'$ existence leads to a non-zero value of this difference, and the measured value of $\sig-\sig^\mathrm{SM}$ is a statistical estimator for $Z'$ parameters.

Let us estimate the statistical uncertainty $\delta\sig$ of the observable (\ref{observ}). Experiment produces events in bins. Let $N_i$ be the number of actual events in the $i$th bin of size $dz_i$. The actual number of events $N_i$ is distributed under the Poisson distribution. This means the mean value of events $\mathbf{M}[N_i]$ coincides with the variance of the number of events $\mathbf{D}[N_i]$,
\begin{equation}\label{dN2a}
\mathbf{D}[N_i]=\mathbf{M}[N_i].
\end{equation}

The number of events can be evaluated by the cross section and the integrated luminosity $\cal L$ of the experiment,
\begin{equation}\label{dNi}
N_i={\cal L}\,\frac{d\sigma_i}{dz_i}\,dz_i.
\end{equation}
Eqs. (\ref{dN2a}) and (\ref{dNi}) give the relation between the mean value and the variance of measured differential cross section:
\begin{equation}\label{dNiD}
\mathbf{D}\left[\frac{d\sigma_i}{dz_i}\right]=\frac{\mathbf{D}[N_i]}{{\cal L}^2\,dz_i^2}=
\frac{1}{{\cal L}\,dz_i}\mathbf{M}\left[\frac{d\sigma_i}{dz_i}\right].
\end{equation}
The obtained relation is a good practical approximation to calculate statistical errors of cross sections in experiments ($\delta\sigma/dz=\sqrt{\mathbf{D}[d\sigma/dz]}$), if we substitute the mean value in the right-hand-side by the actual measured value. For instance, it can be easily checked by existing LEP data on differential cross sections.

If the differential cross section in bins is summed with weights $p_i$, then
\begin{equation}\label{dN}
\mathbf{D}\left[\sum\limits_i p_i\frac{d\sigma_i}{dz_i}dz_i\right]=\sum\limits_i p^2_i dz_i^2\mathbf{D}\left[\frac{d\sigma_i}{dz_i}\right]
=\sum\limits_i  \frac{p^2_i}{{\cal L}}\mathbf{M}\left[\frac{d\sigma_i}{dz_i}\right]dz_i.
\end{equation}
Finally, the last equation can be written in the integral form
\begin{eqnarray}\label{Dint}
&&\mathbf{D}\left[\sig\right]
\simeq
\int\limits_{-1}^{1}dz\,\frac{p^2(z)}{{\cal L}}\,  \frac{d\sigma}{dz},
\end{eqnarray}
where the mean value in the right-hand-side is substituted by the actually measured value

In general, the differential cross section under the integral contains both the contributions from the SM and $Z'$ boson. However, the deviations from the SM are considered to be small. Therefore, in order to simplify calculations we can substitute the cross section by 
its SM part. As a result, the uncertainty of the observable reads
\begin{equation}\label{ds2}
\delta\sig
\simeq \sqrt{\frac{1}{\cal L}\int\limits_{-1}^1 dz\, p^2(z)\frac{d\sigma^\mathrm{SM}}{dz}}.
\end{equation}

In case of massless fermions, the differential cross section of the process $e^+e^-\to\mu^+\mu^-$ with intermediate vector bosons
can be described by only three Legendre polynomials ($P_0$, $P_1$, and $P_2$). Actually, $P_0$ and $P_2$ enters the cross section in the combination $1+z^2$, so a complete orthogonal system for the cross section can be formed by two polynomials. The $Z'$ signal $\sig-\sig^\mathrm{SM}$ also belongs to this system. So, only two polynomials survives in the weight function $p$ in (\ref{observ}). However, they are not the complete system for $p$ when both the mean value and statistical uncertainty are taken into account to determine $p$. The reason is that the statistical uncertainty (\ref{ds2}) contains the squared weight function in contrast to (\ref{observ}). As a result, whatever number of polynomials we try to use to construct $p$ in (\ref{observ}), additional polynomials appear in (\ref{ds2}) by means of the squared weight, and there are no reasons to exclude these additional polynomials in (\ref{observ}). In general, the question about optimal relation between the $Z'$ signal $\sig-\sig^\mathrm{SM}$ and its statistical uncertainty $\delta\sig$ cannot be resolved within a finite system of polynomials. Higher order polynomials can be added to weight function, and, cancelling some part of the cross section by orthogonality, they nevertheless can amplify the measured value with respect to the statistical uncertainty.

We are interested to construct the observable which amplifies the $Z'$ signal as much as possible. This aim can be reached by maximizing the value-to-uncertainty (signal-to-uncertainty) ratio
\begin{eqnarray}\label{max}
&&
\sig-\sig^\mathrm{SM}=
\int\limits_{-1}^{1} dz\,p(z)\left(\frac{d\sigma}{dz}-\frac{d\sigma^\mathrm{SM}}{dz}\right),
\qquad
\delta\sig=
\sqrt{\frac{1}{\cal L}\int\limits_{-1}^{1}dz\, p^2(z)\frac{d\sigma^\mathrm{SM}}{dz}},
\nonumber\\&&\mathrm{abs}\left(\frac{\sig-\sig^\mathrm{SM}}{\delta\sig}\right)\rightarrow\mathrm{max},
\end{eqnarray}
where the weight function is assumed to be varied in the optimization procedure.

As it was mentioned in the previous section, the $Z'$ contribution to the differential cross section (\ref{dsigma}) is approximately 
two-parametric,
\begin{equation}
\frac{d\sigma}{dz}-\frac{d\sigma^\mathrm{SM}}{dz}\simeq
\frac{m^2_{Z}}{4\pi m^2_{Z'}}\left[
F_{1}(z){a}^{2}+F_{4}(z){v}_{e}{v}_{\mu}\right].\label{dsigmasignal}
\end{equation}
Taking proper weight functions $p_a(z)$ or $p_v(z)$, we can suppress one of the factors $F_i(z)$ and obtain one-parametric signals
\begin{eqnarray}\label{oneparamsignals}
&&\sig_a-\sig_a^\mathrm{SM}=\frac{{a}^{2}m^2_{Z}}{4\pi m^2_{Z'}}
\int\limits_{-1}^{1}dz\,p_a(z)F_{1}(z),
\nonumber\\&&
\sig_v-\sig_v^\mathrm{SM}=\frac{{v}_{e}{v}_{\mu}m^2_{Z}}{4\pi m^2_{Z'}}
\int\limits_{-1}^{1}dz\,p_v(z)F_{4}(z) .
\end{eqnarray}
As it is seen, in case of one-parameter observables, the $Z'$ couplings $a$ and $v_{e,\mu}$ in $\sig-\sig^\mathrm{SM}$ are factorized,
and the result of optimization becomes independent of specific values of these unknown $Z'$ couplings. 

In fact, we deal with the constrained optimization. First of all, the normalization of the weight function must be taken into account, since (\ref{max}) is evidently invariant under rescaling of the weight function. We choose the normalization
\begin{eqnarray}
&&\int\limits_{-1}^{1}dz\,p^2\left(z\right)=1.\label{constrN}
\end{eqnarray}

Second, the weight function is chosen to suppress all the factors in the differential cross section (\ref{dsigma}) except for one. Let us choose $F_1$ as an example. The most general scheme takes into account both the contributions of leading factors $F_{1,4}$ and small factors $F_{2,3}$  in the differential cross section (\ref{dsigma}). Of course, two and more factors cannot be integrated to exact zero with single weight function. But we can minimize the cumulative relative contribution of the factors $F_{2,3,4}$ with respect to $F_1$:
\begin{eqnarray}
&&\delta_\mathrm{syst}=\frac{
\sum\limits_{i=2}^4\mathrm{abs}\left(\int\limits_{-1}^{1}dz\,p\left(z\right)F_i\left(\sqrt{s},z\right)\right)
}{
\sum\limits_{i=1}^4\mathrm{abs}\left(\int\limits_{-1}^{1}dz\,p\left(z\right)F_i\left(\sqrt{s},z\right)\right)
}\rightarrow\mathrm{min}.
\label{constrAmin}
\end{eqnarray}
This ratio at the minimum plays the role of the systematic relative error $ \delta_\mathrm{syst}$ of the constructed observable.

Eq. (\ref{constrAmin}) does not specify a unique weight function, it defines a subspace in the Hilbert space of $p(z)$. It is clearly seen from the fact that  (\ref{constrAmin}) does not change when a function orthogonal to $F_{1,2,3,4}$ is added to $p(z)$. As it was mentioned, all the factors $F_{1,2,3,4}$ reside the same two-polynomial subspace of the Hilbert space. So, (\ref{constrAmin}) determines some direction in this two-dimensional subspace.

It worth noticing that the condition (\ref{constrAmin}) can be essentially simplified if the small factors $F_{2,3}$ are neglected. In this case, we obtain the following equation to determines a direction in the two-dimensional subspace corresponding to factors $F_{1,2,3,4}$:
\begin{eqnarray}
&&\int\limits_{-1}^{1}dz\,p\left(z\right)F_4\left(\sqrt{s},z\right)=0.\label{constrA}
\end{eqnarray}
It can be used as a simplified alternative choice to (\ref{constrAmin}). It is also a good start point to numeric optimization (\ref{constrAmin}). In practice, both the conditions (\ref{constrA}) and (\ref{constrAmin}) lead to very close numeric results indistinguishable in plots below.

In case of selecting the term with ${v}_{e}{v}_{\mu}$, the factors $F_1$ and $F_4$ in (\ref{constrAmin}), (\ref{constrA}) switch their roles. As a result, one should minimize the ratio
\begin{eqnarray}
&&\delta_\mathrm{syst}=\frac{
\sum\limits_{i=1}^3\mathrm{abs}\left(\int\limits_{-1}^{1}dz\,p\left(z\right)F_i\left(\sqrt{s},z\right)\right)
}{
\sum\limits_{i=1}^4\mathrm{abs}\left(\int\limits_{-1}^{1}dz\,p\left(z\right)F_i\left(\sqrt{s},z\right)\right)
}\rightarrow\mathrm{min}.
\label{constrVmin}
\end{eqnarray}

The optimization (\ref{max}) with the constraints (\ref{constrN}) and (\ref{constrAmin}) have to determine uniquely the weight function $p(z)$ for the most amplified $Z'$ signal in the considered process.

\section{Kinematics and the polynomials to expand weight functions}
Kinematics of $e^+e^-\to\mu^+\mu^-$ process is relatively simple. Due to the absence of the flavor-changing neutral currents, there is no virtual bosons in the $t$-channel. Moreover, all the leptons can be considered as massless at the ILC energies. This leads to the well-known two-polynomial structure of all the factors in the differential cross sections:
\begin{eqnarray}
&&F_{i}(\sqrt{s},z)  = A_{i}(\sqrt{s})p_{1}(z)+B_{i}(\sqrt{s})p_{2}(z),\label{Fkin}
\end{eqnarray}
where $p_1\sim z$, $p_2\sim (1+z^2)$.
In this regard, it is convenient to use orthogonal polynomials as a basis in the Hilbert space containing all possible weight functions to integrate the cross section. It was mentioned in the previous section that optimized weight functions are not polynomials, so we can construct any convenient basis for numerical computations. We choose $p_1$ and $p_2$ as the first two elements of the orthogonal system of functions. 

We define orthogonal normalized polynomials in the standard way,
\begin{equation}
\int\limits_{-1}^{1}dz\,p_i(z)p_j(z)=\left\{
\begin{array}{cc}
1,& i=j,\\
0,& i\not=j.\\
\end{array}
\right.
\end{equation}

\begin{figure}
\centerline{\includegraphics[scale=0.8]{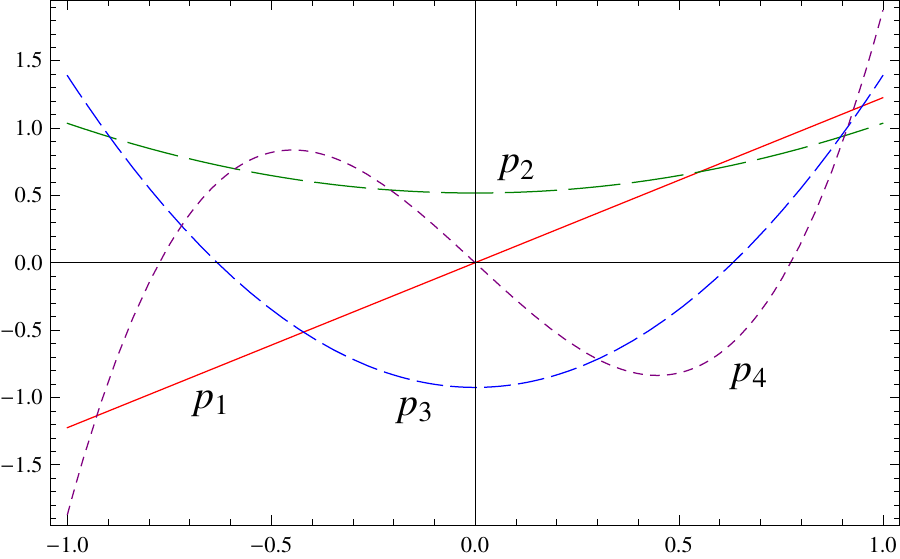}}
\caption{Orthogonal polynomials used as a basis in the Hilbert space of weight functions. \label{fig:pi}}
\end{figure}

We can easily reconstruct the full set of polynomials starting from $p_1$ and $p_2$ and increasing the largest power of the polynomial:
\begin{eqnarray}
&&p_{1}=\sqrt{\frac{3}{2}}z,
\quad
p_{2}=\frac{1}{2}\sqrt{\frac{15}{14}}(z^{2}+1),
\nonumber\\ &&
p_3=\sqrt{\frac{3}{14}}\left(5z^{2}-2\right),
\quad
p_4=\frac{1}{2}\sqrt{\frac{7}{2}}\left(5z^3-3z\right),
\quad
p_5=\frac{3}{8 \sqrt{2}}\left(35 z^4 - 30 z^2 + 3\right),
\nonumber\\ &&
p_6= \frac{1}{8}\sqrt{\frac{11}{2}} \left(63 z^5 - 70 z^3 + 15z\right),
\quad
p_7= \frac{5}{16}\sqrt{\frac{13}{2}} \left(-\frac{231}{5}z^6 + 63 z^4  - 21 z^2 +1\right),
\nonumber\\ &&
p_8= -\frac{1}{16}\sqrt{\frac{15}{2}} \left(429 z^7 - 693 z^5 + 315 z^3 -35z\right),
\quad\ldots
\end{eqnarray}
The first four polynomials are plotted in Fig. \ref{fig:pi}. The first polynomial corresponds to the forward-backward cross section, the second is related to the  total cross section, the third can be interpreted as the center-edge cross section, etc.

Weight function $p(z)$ can be expanded by $p_{i}$:
\begin{equation}\label{wfc}
p(z)=\sum\limits_{i=1}^{\infty}c_{i}p_{i}(z).
\end{equation}
Then, the normalization condition (\ref{constrN}) becomes
\begin{equation}\label{normc}
\sum\limits_{i=1}^{\infty} c_i^2=1.
\end{equation}

In fact, orthogonal polynomials are used just as calculation tool to perform optimization (\ref{max}) to find the most effective weight function. A convenient `natural' basis helps us to obtain results in the most quick and simple way. Let us also note that we can easily update our basis for any symmetric interval $[-z_\mathrm{max},z_\mathrm{max}]$ to take into account the actual angular cuts in experiments.

\section{The optimal observable for axial-vector couplings}

In this section we present the observable containing the sign-definite coupling ${a}^2$.
The signal-to-uncertainty ratio must be maximized to find the weight function $p$. Numerically, the coefficients at polynomial expansion (\ref{wfc}) have to be computed. Thus, we select uniquely the most effective one-parameter observable.

Since the $Z'$ contributions to the cross-section are described by two polynomials $p_{1,2}$, we use the fixed direction in the functional subspace based on $p_{1,2}$ in order to suppress either $F_1$ or $F_4$ factor:
\begin{equation}\label{k}
k=c_{2}/c_{1}.
\end{equation}
In this section we choose the weight function that maximizes the contribution of $F_1$ factor in the observable. This can be done by means of (\ref{constrAmin}). The numerical analysis shows that the corresponding relative weight of $F_1$  is about 0.98. 

There is also the normalization condition (\ref{normc}) allowing to determine one of the coefficients through the others. For instance,
\begin{equation}\label{c1}
c_{1}=\sqrt{\frac{1-c_{3}^{2}-c_{4}^{2}-\ldots}{1+k^{2}}}.
\end{equation}

Thus, two coefficients $c_1$, $c_2$ are explicitly expressed by the other coefficients. As a result, $c_3,c_4,\ldots$ are to be varied to find the maximum (\ref{max}).

\begin{figure}[b]
\centerline{\includegraphics[scale=0.9]{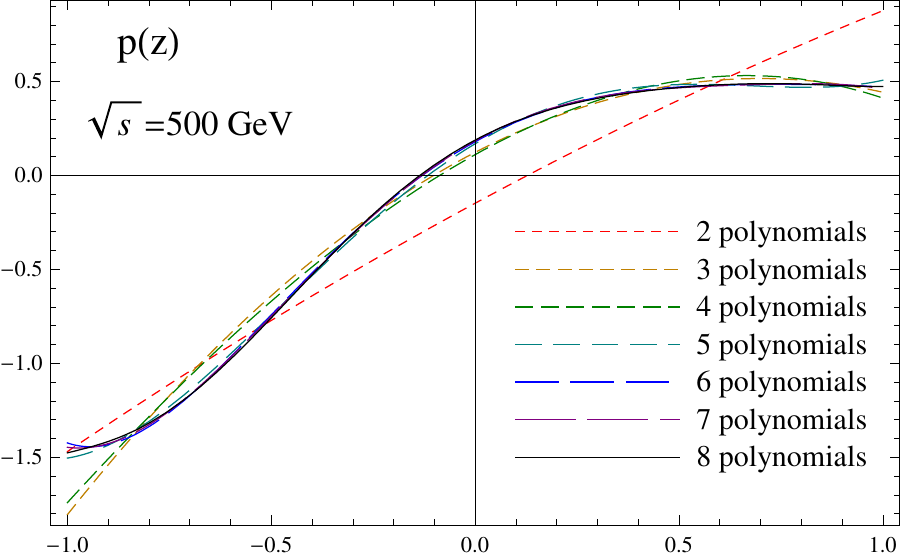}}
\caption{Computing the optimal weight function to select ${a}^2$ at $\sqrt{s}=500$ GeV. \label{fig:pz500}}
\end{figure}

In Fig. \ref{fig:pz500} optimization of the signal-to-uncertainty ratio at $\sqrt{s}=500$ GeV is realized step by step. Increasing the number of polynomials in the weight function expansion we can observe asymptotic behavior of $p(z)$.
We can estimate the relative accuracy of the result comparing the weight functions at the current and previous steps:
\begin{equation}\label{cacc}
\eta=\sqrt{\int\limits_{-1}^{1}dz\,\left(p_\mathrm{current}-p_\mathrm{previous}\right)^2}.
\end{equation}
Using eight polynomials from the basis, we find $\eta<0.01$ at all the considered energies, which is below the theoretical systematic error (2\%) from selecting the sign-definite term in the observable. The results of optimization are shown in Table \ref{tab:opt}.
Numeric value of the signal-to-uncertainty ratio ($R_a$) is defined in the following way:
\begin{equation}\label{Ra}
\mathrm{abs}\left(\frac{\sig-\sig^\mathrm{SM}}{\delta\sig}\right)=\frac{{a}^{2}m^2_{Z}\sqrt{{\cal L}}R_a}{4\pi m^2_{Z'}}.
\end{equation}

\begin{table}[ph]
\centering\caption{The results of optimization of the weight function to select ${a}^2$ or ${v}_e{v}_\mu$ for ILC energies. The parameter $k=c_2/c_1$ is computed in accordance with (\ref{constrAmin}) or (\ref{constrVmin}), then the coefficients $c_i$ in (\ref{wfc}) are found by (\ref{max}). The systematic error of the variable is given by (\ref{constrAmin}) or (\ref{constrVmin}) at the minimum. The optimized numeric value of the signal-to-uncertainty ratio $R_{a,v}$ is defined by (\ref{Ra}) or (\ref{Rv}).}\label{tab:opt}
\begin{tabular}{@{}ccccccc@{}}
\hline\hline  & \multicolumn{3}{c}{axial-vector} & \multicolumn{3}{c}{vector} \\
\hline $\sqrt{s}$, TeV & 0.25 & 0.5 & 1 & 0.25 & 0.5 & 1 \\
\hline 
$k$ & -0.334 & -0.300 & -0.292 & -2.139 & -2.384 & -2.445 \\
$c_1$ & 0.901 & 0.916 & 0.918 & -0.422 & -0.385 & -0.376 \\
$c_2$ & -0.301 & -0.274 & -0.268 &  0.903 & 0.918 & 0.920 \\
$c_3$ & -0.292 & -0.272 & -0.269 & -0.053 & -0.090 & -0.105 \\
$c_4$ & -0.035 & -0.064 & -0.075 & 0.022 & 0.019 & 0.018 \\
$c_5$ & 0.096 & 0.088 & 0.085 & -0.042 & -0.030 & -0.026 \\
$c_6$ & -0.036 & -0.021 & -0.016 & 0.013 & 0.006 & 0.003 \\
$c_7$ & 0.007 & 0.011 & 0.013 & -0.007 & -0.005 & -0.004 \\
$c_8$ & -0.011 & -0.008 & -0.008 & 0.006 & 0.003 & 0.003 \\
$\delta_\mathrm{syst}$, \% & 2.0 & 2.0 & 2.1 & 3.8 & 2.5 & 2.2 \\
$R_{a,v}$, $\times 10^{4} \mathrm{fb}^{1/2}$ & 2.13 & 3.48 & 6.72 & 2.00  & 4.02  & 8.19 \\
\hline \hline 
\end{tabular}
\end{table}

The obtained observable can be physically described as a forward-backward cross section with the smooth step function which is close to non-centered hyperbolic tangent. The forward bins are taken with approximately uniform weights whereas the weight of backward bins increases in the limit $z\to -1$.

\begin{figure}
\centerline{\includegraphics[scale=0.9]{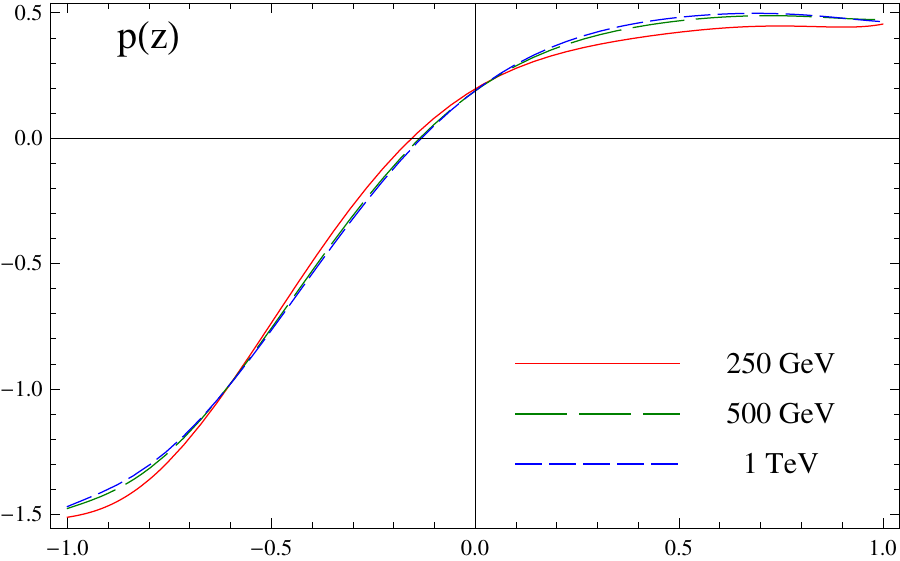}}
\caption{The optimal weight functions to select ${a}^2$ for different ILC energies. \label{fig:pzss}}
\end{figure}

In Fig. \ref{fig:pzss} we show how the optimal weight function depends on the collision energy. As it is seen, the result is stable for different ILC energies. In fact, we can use almost the same function for energies between 0.5 and 1 TeV.

\section{The observable to select vector couplings}

\begin{figure}
\centerline{\includegraphics[scale=0.9]{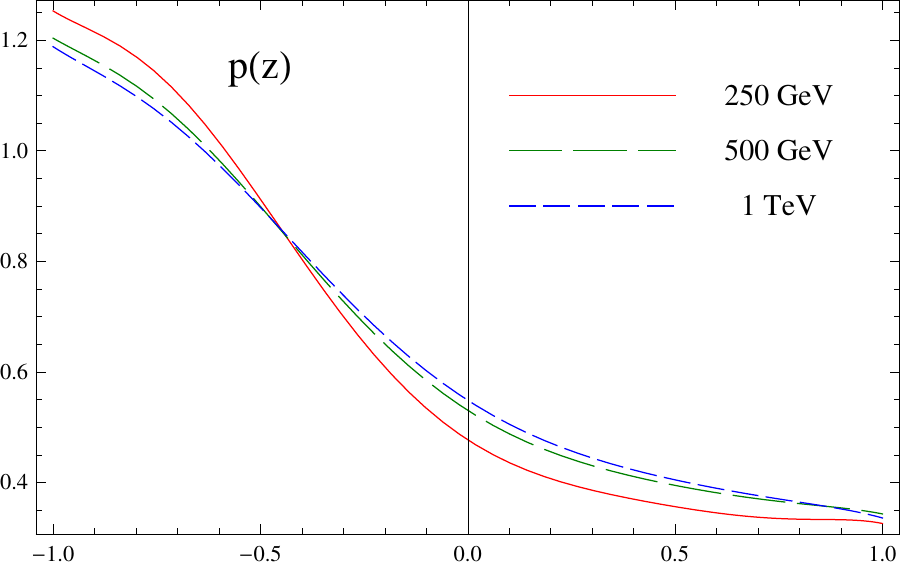}}
\caption{The optimal weight functions to select ${v}_e{v}_\mu$ for different ILC energies. \label{fig:pzssv}}
\end{figure}

The observable to select the term with vector couplings ${v}_e{v}_\mu$ can be constructed in the same way as the observable in the previous section. The only difference is using (\ref{constrVmin}) instead of (\ref{constrAmin}). This gives another direction $k$ in the two-dimensional subspace of the Hilbert space of weight functions. The results of optimization are shown in Table \ref{tab:opt} and Fig. \ref{fig:pzssv}.
Numeric value of the signal-to-uncertainty ratio ($R_v$) is defined in the following way:
\begin{equation}\label{Rv}
\mathrm{abs}\left(\frac{\sig-\sig^\mathrm{SM}}{\delta\sig}\right)=\frac{{v}_{e}{v}_{\mu}m^2_{Z}\sqrt{{\cal L}}R_v}{4\pi m^2_{Z'}}.
\end{equation}

The observable for vector couplings can be also described as a smooth step function. However, all the scattering angles are integrated with the same sign in contrast to the observable for the axial-vector coupling. The forward bins are taken with approximately uniform weights again, whereas the weight of backward bins increases in the limit $z\to -1$.

\begin{figure}
\centerline{\includegraphics[scale=0.6]{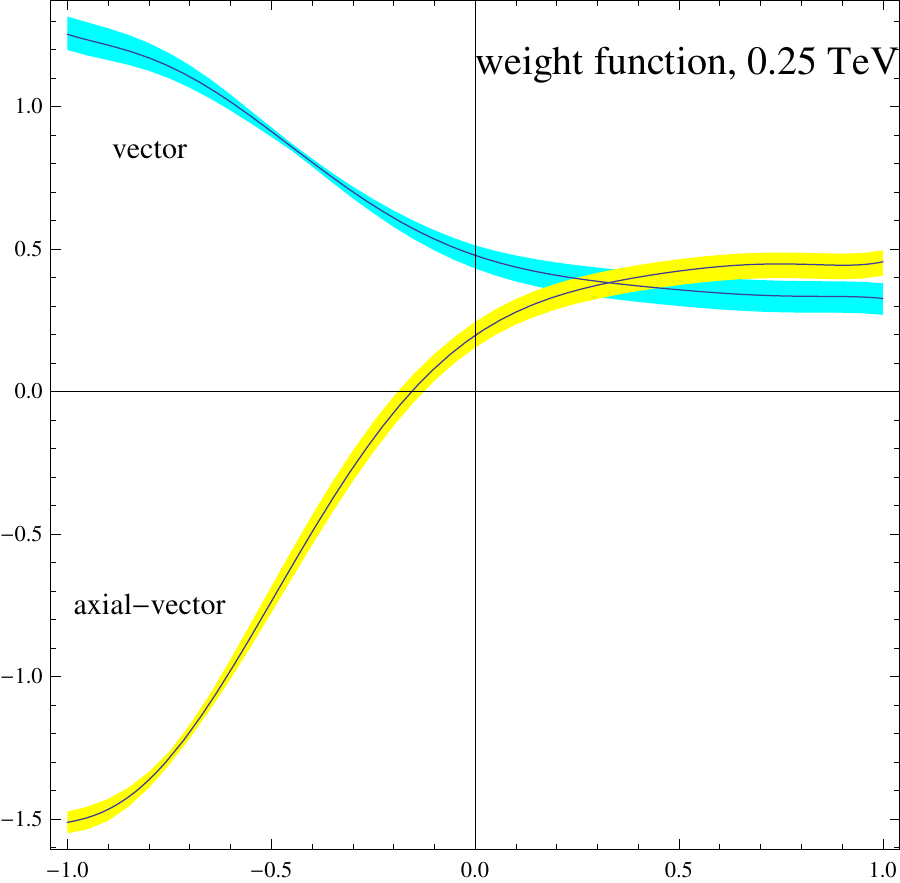}~\includegraphics[scale=0.6]{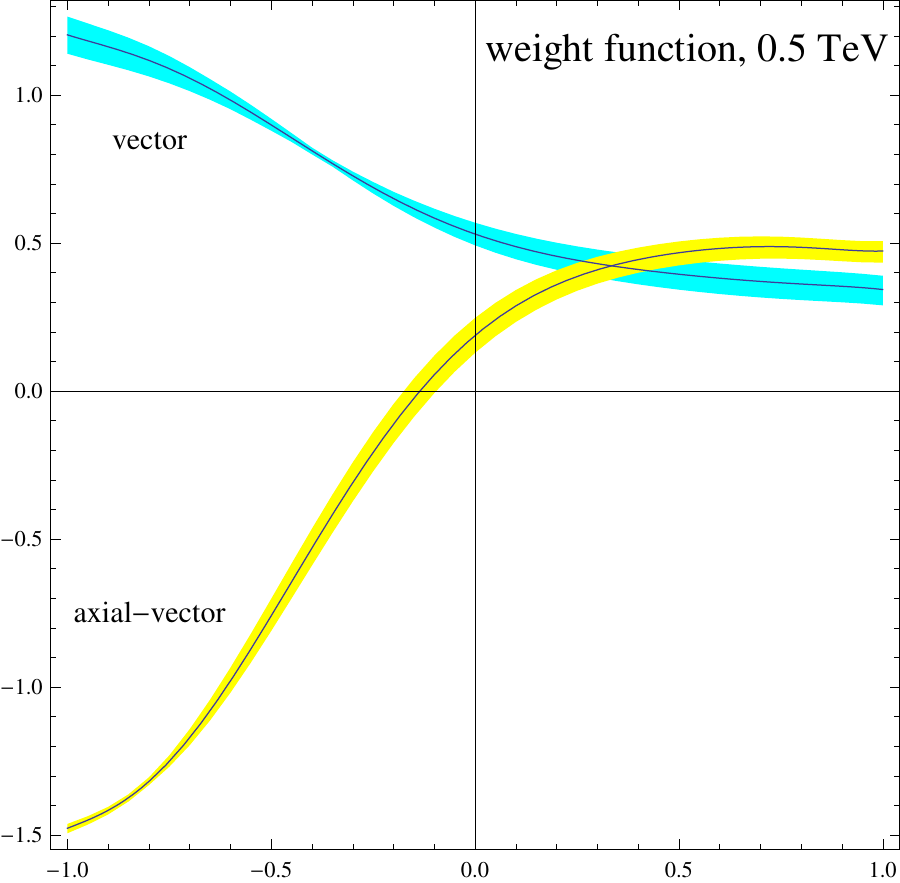}}
\caption{The analysis of possible systematic errors of the weight functions at $\sqrt{s}=250, 500$ GeV. \label{fig:pzrad}}
\end{figure}

\begin{figure}
\centerline{\includegraphics[scale=0.6]{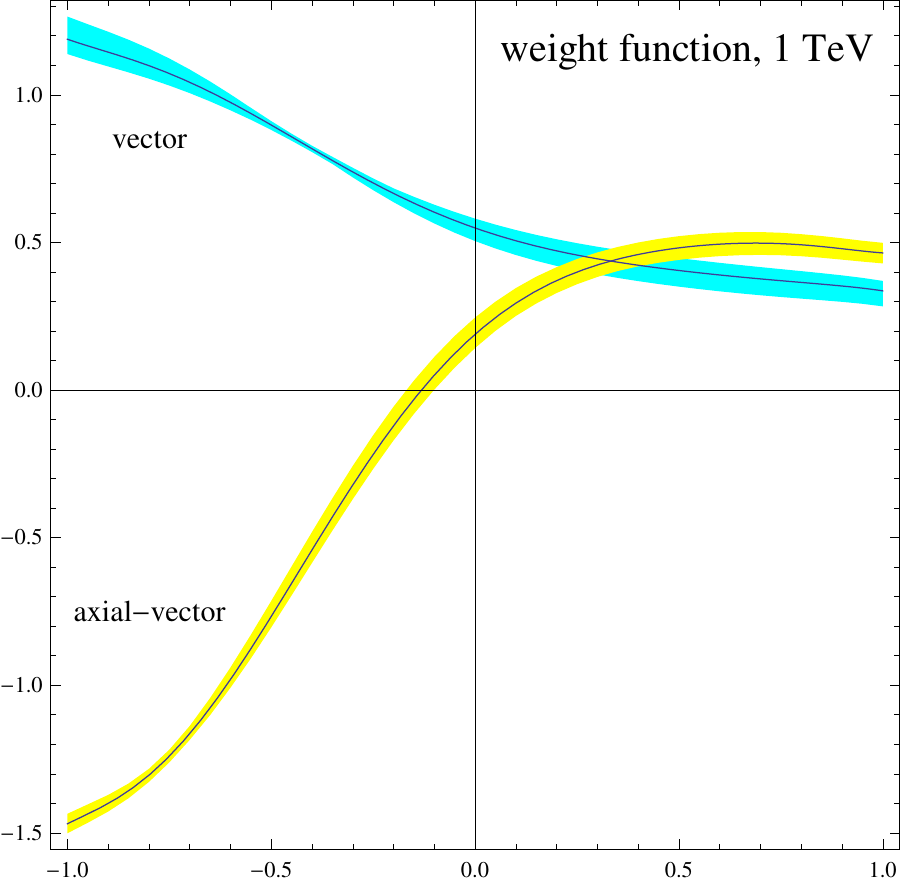}}
\caption{The analysis of possible systematic errors of the weight functions at $\sqrt{s}=1$ TeV. \label{fig:pzrad1}}
\end{figure}

In Figs. \ref{fig:pzrad}--\ref{fig:pzrad1} we show the role of possible systematic errors of cross sections used in our analysis. The shaded areas arise from systematic errors on factors $F_i$ of order about $\pm 5\%$ and also from the systematic errors on the SM cross section of order about $\pm 2\%$ used to calculate statistical uncertainty of observables. It is seen, the weight functions are stable with respect to systematic errors. In principle, systematic errors can be reduced in future data fits by taking into account parameters and settings of actual detectors as well as by more precise calculations.

\section{Discussion}

We propose one-parameter integrated cross sections for $e^{+}e^{-}\to\mu^{+}\mu^{-}$ with the best signal-to-uncertainty ratio.
They allow to select either vector or axial-vector $Z'$ couplings. The observable for axial-vector coupling is sign-definite due to the relations between the couplings of the Abelian $Z'$ boson. In case of lepton universality, the observable to select vector $Z'$ couplings also becomes sign definite.
The observables can amplify possible $Z'$ signals comparing to the other existing observables. In this regard, they can be useful in future experiments at lepton colliders such as the ILC. 

Being evaluated at the maximum, the optimization criterion (\ref{max}) gives an estimate for the $Z'$ signal measured in standard deviations (so called `sigmas'). Of course, to derive a numeric value, we have to suppose some values of the integrated luminosity as well as the unknown common factor in the numerator containing the $Z'$ couplings (either ${a}^2$ or ${v}_e{v}_\mu$ depending on the type of the observable) and the $Z'$ mass. Nevertheless, it is interesting to obtain some estimates taking into account either current bounds on $Z'$ couplings or popular model settings.

Before the calculation of possible $Z'$ signals, we must say a couple of words about systematic errors of experimental data. Although in our analysis the theoretical prediction for the $Z'$ signal $\sig-\sig^\mathrm{SM}$ can be computed directly by the right-hand-side of (\ref{dsigma}), the actual experimental value will be computed by the left-hand-side of (\ref{dsigma}). Thus, the systematic errors are possible in measured signal, since the SM value is always defined up to some theoretical uncertainties and backgrounds. It is better to include to the SM value of the cross-section important backgrounds which might be dangerous for extracting
small off-shell $Z'$ signal.  For example, one can mention four-fermion process $e^+e^- \to e^+e^-\mu^+\mu^-$ with the electron lost in the beam-pipe. Nevertheless, some of backgrounds always remain. They can be included into consideration as systematic error of measured signal taken often as 0.2\% of the SM value \cite{Baer:2013cma}. We also use $\sig^\mathrm{syst}=0.002\sig^\mathrm{SM}$. Below, the systematic errors will be given in units of statistical error, $\sig^\mathrm{syst}/\delta\sig$, in order to make the comparison with the signal-to-uncertainty ratio $(\sig-\sig^\mathrm{SM})/\delta\sig$.

The ILC `canonical program' consists of runs at $\sqrt{s}=$ 0.25, 0.5, and 1 TeV with integrated luminosities ${\cal L}=$ 250, 500, and 1000 fb${}^{-1}$, respectively \cite{Baer:2013cma}. As for other unknown quantities, let us compare estimates of $Z'$ couplings in our approach with the analysis of particular $Z'$ models. In \cite{Baer:2013cma} (Fig. 3.2) one can find confidence areas in $a_l$-$v_l$ plane for the $\chi$ model with $m_{Z'}=$ 3 TeV, $v_l=2a_l\simeq 0.4$. With this settings, taking ${\cal L}=500$  fb${}^{-1}$ at $\sqrt{s}=0.5$ TeV and ${\cal L}=1000$ fb${}^{-1}$ at $\sqrt{s}=1$ TeV we can obtain from (\ref{Ra}), (\ref{Rv}) signal-to-uncertainty ratios. These values can be interpreted as the signal strength in numbers of standard deviations (`sigmas'). For the `axial-vector' observable, we obtain $3.5\sigma$ (0.5 TeV) and $6.7\sigma$ (1 TeV) signals, whereas the `vector' observable leads to $10.5\sigma$ (0.5 TeV) and $30.3\sigma$ (1 TeV) signals. At the same time, the mentioned 0.2\% systematic errors are one order less than the signal being $0.27\sigma$ (0.5 TeV, axial-vector) $0.19\sigma$ (1 TeV, axial-vector), $0.81\sigma$ (0.5 TeV, vector), and $0.57\sigma$ (1 TeV, vector). Thus, the SM prediction of the cross section must be computed with the accuracy better than 1\% to see such $Z'$ signals.

The considered cross-sections contain mainly the spin-one exchange amplitude. In case of polarized bins, spin-zero exchange can be suppressed in measured data giving amplification of the considered cross-sections up to two times. In accordance with (\ref{max}), further amplification of the signal-to-uncertainty ratio up to $\sqrt{2}$ times can be supposed dependently on actual polarization of incoming bins. Fig. 3.2 in \cite{Baer:2013cma} shows also the possibility of measuring $Z'$ couplings in the $\chi$ model with $m_{Z'}=$ 3 TeV by means of two-parameter fit with highly polarized incoming bins ($P_{e-}=0.8$ and $P_{e+}=0.6$). One can see approximately $4.5\sigma$ (0.5 TeV) and $12\sigma$ (1 TeV) signals for the axial-vector coupling, $6\sigma$ (0.5 TeV) and $23\sigma$ (1 TeV) signals for the vector couplings. Taking into account that such polarized bins amplify spin-one exchange cross-section by factor $(1+P_{e-})(1+P_{e+})/2=1.44$, we can multiply our results above by $\sqrt{1.44}\simeq 1.2$ and conclude that our approach has prominent experimental perspectives comparing to standard model-dependent analysis. To compare  smaller $Z'$ couplings, we can use Fig. 3.3 in \cite{Baer:2013cma} corresponding to $m_{Z'}=$ 4 TeV, $\sqrt{s}=$ 0.5 TeV,  ${\cal L}=$ 1000 fb${}^{-1}$ and polarized incoming bins. In this case we still obtain for the $\chi$ model the signals of $1.9\sigma$ (axial-vector) and $8.7\sigma$ (vector) even without accounting for polarization, whereas one can see no significant signals in the two-parametric fit. The 0.2\% systematic errors for these settings become $0.39\sigma$ (axial-vector) and $1.15\sigma$ (vector) being below the signal level.

 It is worth noticing that the settings above are quite `optimistic' for $Z'$ discovery at the ILC. With these values of the $Z'$ couplings and mass, the four-fermion couplings $a^2m_Z^2/(4\pi m_{Z'}^2)\simeq 3\times 10^{-6}$, $v^2m_Z^2/(4\pi m_{Z'}^2)\simeq 1.2\times 10^{-5}$. 
Although the LEP data constrain the four-fermion couplings by $10^{-5}$ (see \cite{Gulov:2010zq}), the latest LHC searches for $Z'$ resonance constrict the upper bound of the couplings rather to $10^{-6}$ (see \cite{Gulov:2013ijmpa}). These values could correspond to the non-leptophobic $Z'$ with the mass more than 3 TeV.  Thus, we can perform `pessimistic' estimate (supposing either weaker $Z'$ couplings or heavier $Z'$ mass), substituting ${a}^2 m_Z^2/(4\pi m_{Z'}^2)$ or ${v}_e{v}_\mu m_Z^2/(4\pi m_{Z'}^2)$ in the numerators of (\ref{Ra}), (\ref{Rv}) by $10^{-6}$ without further specification of the couplings $a$, $v_{e,\mu}$ and the $Z'$ mass.  Assuming the ILC integrated luminosities mentioned above, we could obtain the $Z'$ signals at $0.3\sigma$, 0.8$\sigma$, and 2.1--2.6$\sigma$ at $\sqrt{s}=$ 0.25, 0.5, and 1 TeV. Thus, even in this `pessimistic' case, the $Z'$ signal can be observed at 3-4$\sigma$ level in the combined fitting of all the ILC runs. 

The ideas of the present paper can be applied to processes with more complicated kinematics.


\end{document}